\documentclass[twocolumn,showpacs,superscriptaddress,prb]{revtex4-1}
\usepackage{graphicx}
\usepackage{color}
\usepackage{amsmath,amsthm,amssymb}
\usepackage{braket}
\usepackage[colorlinks,citecolor=blue,linkcolor=red]{hyperref}

\usepackage{soul}
\begin{document}
 
\title{Thickness dependence of electronic structure and optical properties of a correlated van der Waals antiferromagnet NiPS$_3$ thin film}

\author{Christopher Lane}
\email{laneca@lanl.gov}
\affiliation{Theoretical Division, Los Alamos National Laboratory, Los Alamos, New Mexico 87545, USA}
\affiliation{Center for Integrated Nanotechnologies, Los Alamos National Laboratory, Los Alamos, New Mexico 87545, USA}

\author{Jian-Xin Zhu}
\email{jxzhu@lanl.gov}
\affiliation{Theoretical Division, Los Alamos National Laboratory, Los Alamos, New Mexico 87545, USA}
\affiliation{Center for Integrated Nanotechnologies, Los Alamos National Laboratory, Los Alamos, New Mexico 87545, USA}

\date{version of \today} 
\begin{abstract}
We study the thickness dependence of the electronic, magnetic, and optical properties of a NiPS$_3$ thin film, an antiferromagnetic charge-transfer insulator. Utilizing state-of-the-art advanced density functionals, we find the antiferromagnetic Zig-Zag order, the band gap, and the main peaks in the dielectric tensor are all in good agreement with the corresponding experimental values. Upon thinning, the Zig-Zag antiferromagnetic order becomes virtually degenerate with a competing N\'eel order, consistent with the suppression of long-range order observed by Raman spectroscopy due to strong magnetic fluctuations. Additionally, due to the robustness of the electronic band gap observed by spectroscopic ellipsometry measurements above $T_N$, we suggest that the persistence of the band gap is driven by strong electronic correlations. Other systematic changes in electronic dispersion, effective mass, and Kerr angle with thickness are also discussed. Finally, an applied external electric field is found to suppress the band gap by up to 13\%, until precipitating an insulator-metal transition at a critical field value of $0.7$ eV/\AA. 
\end{abstract}


\maketitle 
\section{Introduction}
Since the isolation of graphene in 2008, there has been an explosion in the number of two-dimensional (2D) atomically-thin films predicted and synthesized using atomic species from across the periodic table. These 2D thin-films exhibit a wide variety of emergent novel phases including electronic,~\cite{kim2015observation} excitonic,~\cite{pollmann2015resonant} valley,~\cite{rivera2016valley} and correlated physics~\cite{costanzo2016gate} all under 2D confinement. Owing to their varied properties, 2D materials hold immense potential in a diverse spectrum of technological applications including optoelectronics,~\cite{cheng2019recent}  single-molecule detection,~\cite{feng2015identification} and energy storage and harvesting.~\cite{han2017recent,hennighausen2019oxygen,lane2019understanding}

The discovery of 2D materials exhibiting strong magnetic correlations and long-range order is nontrivial and has garnered substantial attention not only for their practical uses in advancing spintronics and quantum information sciences, but also for the fundamental questions they raise. Due to the 2D confinement, exotic quantum phases of matter have been predicted including 2D Kitaev spin liquids~\cite{takagi2019concept} and fractionalized charge states,~\cite{castelnovo2012spin} along with new elementary excitations such as Majorana fermions. Though a few candidate materials have been found, none have been confirmed for these exotic states.~\cite{takagi2019concept} Additionally, the high-temperature superconducting cuprate and iron-pnictide material families are composed of antiferromagnetic (AFM) 2D CuO$_2$ or Fe$Pn$ ($Pn$=As, P) planes interwoven with spacer layers comprised of rare-earth elements.~\cite{kastner1998magnetic,wen2011materials} These 2D planes exhibit unconventional superconductivity and a myriad of other stripe, charge density wave, and correlated metal phases. Therefore, the discovery of other 2D materials displaying similar phase diagrams may provide valuable insights into the mechanism of unconventional superconductivity and the pseudogap regime.

In particular, the $M$PS$_3$ ($M$=Mn, Fe, Co, Ni) family of compounds are quasi-2D with their crystal structure analogous to graphene.~\cite{brec1986review}  The transition metal $M$ is octahedrally coordinated by sulfur ligands within a single layer. The S atoms are connected to two P atoms located above and below the Ni plane producing a honeycomb lattice in the $ab$-plane.  The atomically-thin films are then stacked in the $c$-direction in an $AB$ manner.  Since two P atoms and six S atoms are covalently bonded among themselves, forming a (P$_{2}$S$_{6}$)$^{4-}$ anion complex, each transition metal carries a $2^+$ ionic state.\cite{wang2018new}  Therefore, all members of the family display various long-range magnetic orders including ferromagnetic (FM), Zig-Zag, N\'eel, and stripy AFM  on the transition metal sites. Moreover, the magnetic configurations follow the Ising, XY, and Heisenberg spin models, making this family of materials a unique platform for studying the phase diagram of these models under pressure, doping, and external fields.

NiPS$_3$ has recently gained attention for following the highly anisotropic XXZ Heisenberg model, where upon thinning of the film, magnetic fluctuations are found to dominate, suppressing the emergence of long-range order down to very low temperatures, in accord with the Kosterlitz-Thouless (KT) 
transition.~\cite{kim2019suppression,kosterlitz1973ordering,kim2019suppression} 
A possible Mott metal-insulator transition under pressure~\cite{kim2018mott} similar to that of the high-$T_c$ cuprates is also predicted.
Furthermore, an optical spectroscopy study of NiPS$_3$ reports three predominent transitions in the optical conductivity and their evolution with temperature.~\cite{kim2018charge} 
Although NiPS$_3$ has gained interest among the experimental community, there is currently no comprehensive study, which is devoted to examining the electronic and optical properties.

In this article, we present a first-principles investigation of ultra-thin films of NiPS$_3$, and systematically analyze the evolution of the ground state magnetic configuration and electronic structure with film thickness. The Zig-Zag AFM order is found to be the ground state for all values of the thickness. However, for a single layer, the Zig-Zag and N\'eel orders are virtually degenerate ($\delta E\sim0.2$ meV/Ni) facilitating large magnetic fluctuations, consistent with the suppression of magnetic ordering observed by a recent Raman spectroscopy study.~\cite{kim2019suppression}  Furthermore, we calculate the frequency dependent dielectric tensor for various film thicknesses and compare to the measured values. The theoretically obtained three prominent peaks along with their line shape are in good agreement with the observed optical spectra.\cite{kim2018charge} Moreover, a weak but significant magneto-optical response is found, where the leading edge of the complex Kerr parameters follow a systematic trend with thickness and magnetic configuration. Additionally, due to the extreme flatness of the conduction and valence bands, we find the electronic carriers to be quite heavy, exhibiting effective masses of 9 - 10 $m_e$ on average. Finally, we find an applied external electric field along the $z$-axis to tune down the band gap by up to 13\%, until destabilizing the AFM order at a critical field of $0.7$ eV/\AA~ causing an insulator-metal transition.

The outline of this paper is as follows. In Sec.~\ref{sec:compdetails} the computational details are summarized.  In Sec.~\ref{sec:mag} the crystal structure is introduced along with a comparison of the various magnetic orderings. In Sec.~\ref{sec:elecoptical}
the electronic structure, effective masses, and dielectric function of the Zig-Zag AFM ground state are presented. Sec.~\ref{sec:efield} discusses the effect of an external electric field on the electronic and magnetic state. Finally, Sec.~\ref{sec:conclusion} is devoted to the conclusions.

\section{Computational Method}\label{sec:compdetails}

{\it Ab initio} calculations were carried out by using the pseudopotential projector-augmented wave method\cite{Kresse1999} implemented in the Vienna ab initio simulation package (VASP) \cite{Kresse1996,Kresse1993} with an energy cutoff of $300$ eV for the plane-wave basis set. Exchange-correlation effects were treated using the strongly constrained and appropriately normed (SCAN) meta-GGA scheme.~\cite{Sun2015} A 8 $\times$ 4 $\times$ 1  (8 $\times$ 4 $\times$ 8) $\Gamma$-centered k-point mesh was used to sample the slab (bulk) Brillouin zone. A denser mesh of 12 $\times$ 6 $\times$ 1  (12 $\times$ 6 $\times$ 12) was employed for the calculation of the dielectric tensor. Spin-orbit coupling effects were included self-consistently. The experimentally obtained atomic positions and lattice paramters for the bulk, trilayer, and bilayer C2/m (Space group number: 12) structure and the hexogonal D$_{3d}$ monolayer were used throughout this work.~\cite{taylor1973preparation} \footnote{ 
The performance of SCAN within the 2D van der Waals class of materials has been addressed by Buda {\it et al.}\cite{buda2017characterization}, where SCAN yields significant improvements over LSDA, PBE, and PBEsol. Additionally, SCAN has been shown to accurately predict the crystal structure and many other key properties of the transition-metal oxide compounds \cite{Furness2018,Lane2018,zhang2020competing,lane2020first} as compared to standard and hybrid density functional approximations \cite{pokharel2020ab}.  } A total energy tolerance of $10^{-6}$ eV was used to determine the self-consistent charge density.  Because the experimental  Neel temperature is as high as $T_N \sim158$ K, all of our calculations are focused on the magnetic state.

\section{Magnetic Ground State}\label{sec:mag}

\begin{figure}[ht!]
\includegraphics[width=0.97\columnwidth]{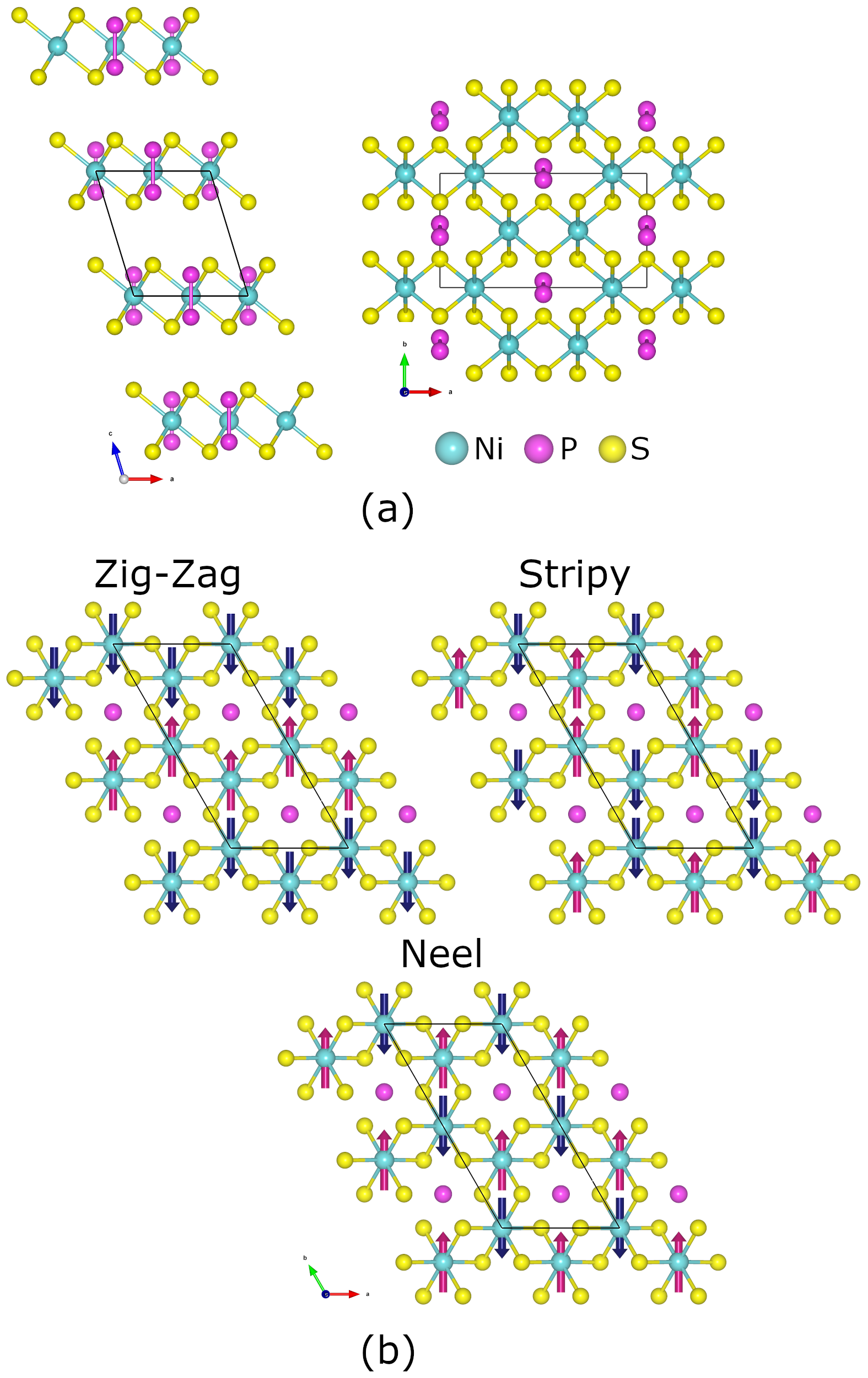}
\caption{(color online) (a) The crystal structure of bulk NiPS$_3$ seen along the $c$ and $b$ axes. The monoclinic symmetry of the bulk structure is exhibited by the AB stacking of the atomic layers along the $c$-axis. (b) Various antiferromagnetic ground state configurations within a monolayer of NiPS$_{3}$.  Red (Blue) arrows represent the positive (negative) nickel magnetic moments. The black lines mark the unit cell. } 
\label{fig:MagArrangements}
\end{figure}

\begin{figure*}[ht!]
\includegraphics[width=1.0\textwidth]{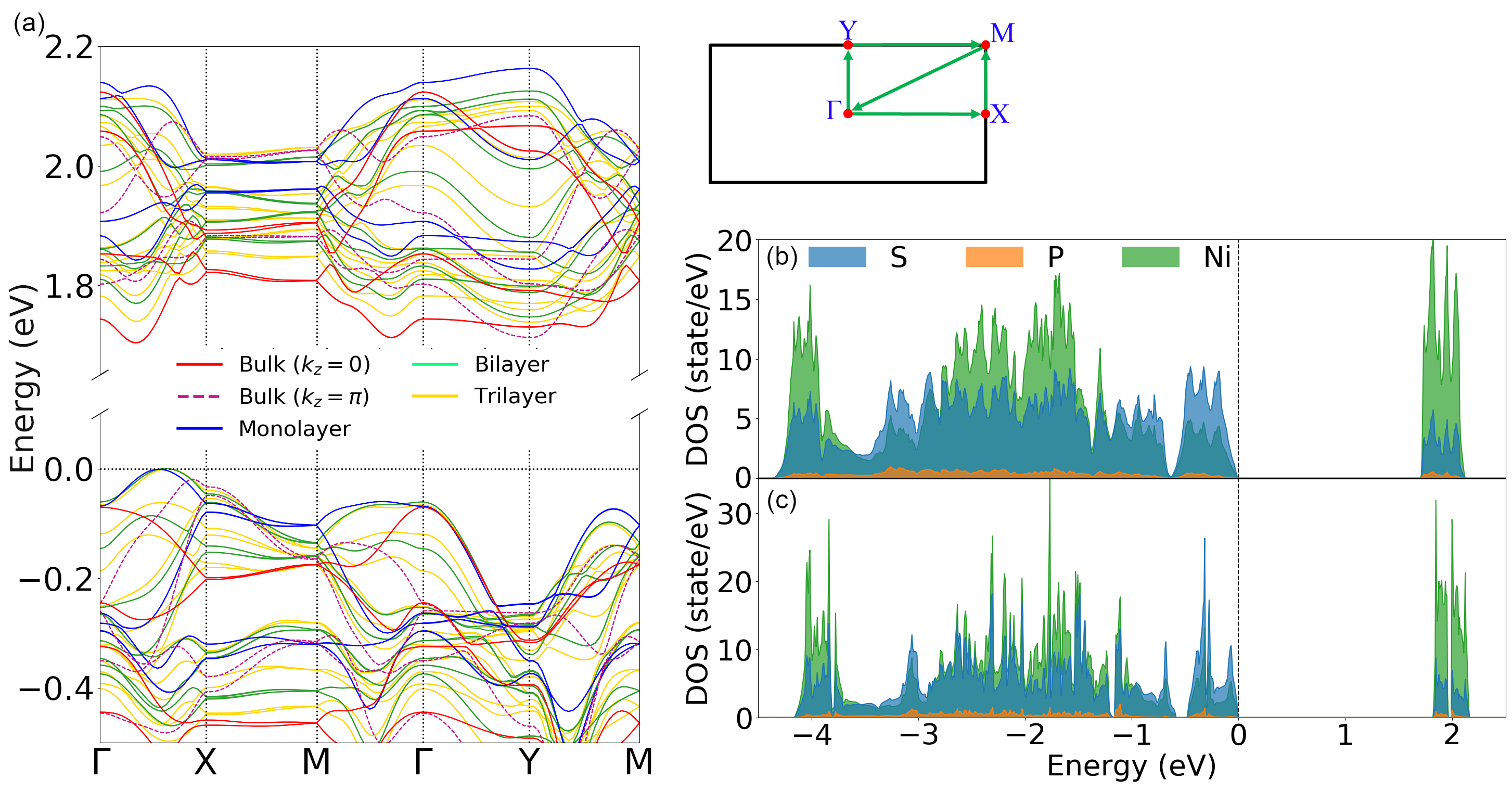}
\caption{(color online)  (a) Electronic band dispersion of NiPS$_3$ in the Zig-Zag AFM phase for various thicknesses. A schematic of the Brillouin zone; where the path followed in the electronic dispersions in (a) is marked.  (b) and (c) shows the atomic-site projected partial density-of-states of bulk and monolayer Zig-Zag AFM NiPS$_3$, respectively.   } 
\label{fig:BANDSDOS}
\end{figure*}

Figure~\ref{fig:MagArrangements}(a) shows NiPS$_3$ in the bulk monoclinic structure, where the nickel atoms (teal spheres) are arranged in a honeycomb pattern and are octahedrally coordinated by six sulfur atoms (yellow spheres). The sulfur atoms are also bonded to two phosphorous atoms (pink spheres) which sit above and below the nickel layer. The NiPS$_3$  layers are stacked along the $c$-axis such that a sulfur atom of one layer sits directly above the phosphorous of the next, resulting in a relative shift of the unit cell along the $a$-axis. The layers are weakly bound to each other by van der Waals interactions, enabling the exfoliation of atomically thin few-layer samples. Upon thinning, the monoclinic crystal symmetry is maintained until the monolayer, where the point group changes to the hexagonal D$_{3d}$.

Figure~\ref{fig:MagArrangements}(b) shows the three low-energy magnetic configurations of NiPS$_3$ within the generalized honeycomb crystal structure of the monolayer. Magnetic moments are stabilized on the nickel atomic sites (red and blue arrows) oriented within the $ab$-plane, perpendicular to the $a$-axis, in accord with recent neutron diffraction results.\cite{wildes2015magnetic} Our calculations show that the $15^{\circ}$ out-of-plane spin tilt is energetically similar ($\sim 1$ meV) from the $0^{\circ}$ orientation in correspondence with the relatively weak spin-orbit coupling found in atomic nickel. The predicted value of the magnetic moment on nickel sites in the Zig-Zag phase is $1.504 \mu_{B}$ which is in agreement with the average experimental value of  $1.12 \mu_{B}$.~\cite{wildes2015magnetic} Our nickel magnetic moments are slightly enhanced compared to the experimental value  due to the oversensitivity of the so-called iso-orbital indicator ($\alpha$) used in SCAN to distinguish between various chemical bonding environments. Improvements in the SCAN functional in this connection, \cite{mejia2019analysis,furness2019enhancing} however are not likely to to significantly change the conclusions of the present study.  Curiously, this oversensitivity was not found in studies of the high temperature superconducting cuprates,~\cite{Furness2018,Lane2018,zhang2020competing} and 3$d$ and 5$d$  perovskite oxides in general.~\cite{lane2020first,varignon2019origin,zhang2019symmetry}

\begin{table}[ht!]
\caption{\label{table:groundstate}Comparison of various theoretically predicted  properties for the three possible antiferromagnetic ground states for  bulk and monolayer NiPS$_3$. The AFM orders presented for the bulk case follow the C-type stacking along the $c$-axis.} 
\begin{ruledtabular}
\begin{tabular}{cccccc}
Order &  Spin & Orbital & Total & Gap & Relative Energy\\
& ($\mu_{B}$) & ($\mu_{B}$) & ($\mu_{B}$) &  (eV) & (meV/Ni)\\
\hline\hline
Bulk\\
Zig-Zag & 1.434 & 0.069 & 1.504 & 1.719 & 0\\
N\'eel       & 1.438 & 0.069 & 1.508 & 1.623 & 1.12\\
Stripy     & 1.474 & 0.071  & 1.547 & 1.027 & 52.81\\
\hline
Monolayer\\
Zig-Zag & 1.435 & 0.069 & 1.504 & 1.827 & 0\\
N\'eel       & 1.438 & 0.069 & 1.508 & 1.905 & 0.19\\
Stripy     & 1.475 & 0.071  & 1.546 & 1.169 & 53.83\\
\end{tabular}
\end{ruledtabular}
\end{table}

\begin{table*}[ht!]
\caption{\label{table:effmass}The effective mass at the conduction and valence band edges, calculated using three point forward finite difference equation in units of $m_e$. The band edge extrema curvature are evaluated along various directions ($x$,$y$,$diagonal$) in the Brillouin zone as indicated by the subscript. The sign indicates whether the effective mass is hole ($+$) or electron ($-$) like. The bands that display a linear dispersion are marked with a dash (-).   } 
\begin{ruledtabular}
\begin{tabular}{ccccccccccc}
& $\Gamma_x$ & $\Gamma_y$ & $\Gamma_{diagonal}$ & $\Gamma -X$ & $X_y$ & $Y_y$ & $Y_x$ & $M_y$ & $M-Y$ & $M-\Gamma$ \\
\hline\hline
Valence\\
Monolayer        & 1.85 & -0.67 &  10.80 & -3.11 & -2.21 & -5.91 & 2.01 & -13.17 & -2.22 & -0.83  \\
Bilayer          & 3.09 & -0.70 & -13.01 & -3.72 & -7.72 & -4.41 & 0.74 &   2.22 & -2.15 & -11.54 \\
Trilayer         & 3.66 & -0.72 &  -9.26 & -3.39 & -1.14 & -6.32 & 1.22 &   1.90 & -1.92 & -11.54 \\
Bulk $(k_z=0)$   &-9.59 & -0.72 &  -2.55 & -     &  8.65 &  0.51 & 2.02 &  -6.10 & -3.14 &	4.00 \\
Bulk $(k_z=\pi)$ &-     & -0.71 &-       & -1.12 & -1.00 & 27.12 & 2.60 &   2.03 & -2.94 &  -6.18 \\
\hline
Conduction\\
Monolayer        &-1.57 & -26.04 & -1.92 &  1.50 & 54.54 &  3.39 & 4.25 &  -7.79 &  2.19 &   2.02 \\
Bilayer          &-1.57 & -22.14 & -1.97 &  1.44 & -7.72 &  2.97 & 5.52 & -11.93 &  2.02 &   2.15 \\
Trilayer         &-1.51 & -23.98 & -1.95 &  1.41 & -5.92 &  2.73 & 7.70 &  -9.88 &  1.96 &   2.18 \\
Bulk $(k_z=0)$   &-1.57 & -14.45 & -2.05 &  1.42 & -3.94 &  8.33 & 6.25 & -14.54 &  4.03 &   2.05 \\
Bulk $(k_z=\pi)$ & 6.56 &  -1.21 & -0.57 &  1.54 & -148.85& 2.57 & 1.82 & -53.34 &  5.13 &   3.24 \\
\end{tabular}
\end{ruledtabular}
\end{table*}

Our {\it ab initio} total energy calculations find the Zig-Zag type AFM order to be the ground state for all values of the film thickness in agreement with experimental observations. \cite{kim2018charge,wildes2015magnetic} For the bulk crystal,  the N\'eel and Stripy phases are found to lie at $1.12$ meV and $52.81$ meV above the ground state, respectively. Moreover, we also find the C-type AFM order along the $c$-axis to be marginally ($\sim 1$ meV/Ni) more stable as compared to G-AMF, indicative of weakly bound van der Waals materials.   Interestingly, as the number of layers of NiPS$_3$ is decreased below five layers, the relative  energetic separation between Zig-Zag and N\'eel orderings decreases by an order of magnitude, resulting in a near degeneracy. The large fluctuations produced by the effective degeneracy of Zig-Zag and N\'eel configurations is consistent with the suppression of magnetic ordering observed in Raman spectroscopy.~\cite{kim2019suppression} This behavior resembles the pseudogap regime in the underdoped cuprates where competition between various magnetic ordering states suppress ordering down to low temperatures.~\cite{markiewicz2017entropic} Furthermore, a recent ARPES study of FeSe suggests Kosterlitz-Thouless physics may play a key role in the pseudogap regime, similar to NiPS$_3$.~\cite{shen2020}  Table~\ref{table:groundstate} gives the magnitude of the magnetic moments along with the band gap and relative total energy of the various magnetic configurations.

\section{Electronic Structure and Optical properties}\label{sec:elecoptical}

Figure~\ref{fig:BANDSDOS} shows the electronic band structure and  atomic-site projected partial density-of-states for various values of film thickness of NiPS$_3$ in the Zig-Zag AFM phase. An AFM state stabilizes over the nickel atoms by splitting the up- and down-spin of the Ni-S antibonding states producing a gap in the Ni-$d$ bands. As a result of strong electron-electron interactions, the nickel dominated (82\%) conduction states appear `mirrored' at -4 eV, bookending the full bandwidth of Ni-S hybridized levels. In between, the valence states are mainly composed of sulfur-$p$ orbitals, accounting for 66\% of the total atomic weight. This stacking-of-states follows the Zaanen-Sawatzky-Allen classification of a charge-transfer insulator.~\cite{zaanen1985band} Therefore, in contrast to the Mott-Hubbard insulator, when a hole is doped into NiPS$_3$, the carrier would sit on the sulfur atomic sites rather than in the nickel sites.

The AFM order produces a 1.827 eV band gap in the monolayer case. The gap is indirect, with the lowest energetic transition occurring at $(\frac{\pi}{2},0)$ and $(\frac{\pi}{2}\mp\delta k,0)$, for the valence and conduction bands, respectively. However, only a small phonon-assisted momentum transfer of $\delta k\approx \pm 0.017\cdot\frac{\pi}{2}$  is needed to connect valance and conduction band edges, allowing for efficient optical absorption and emission. Due to very weak  spin-orbit coupling exhibited by the relatively light nickel atoms, spin splitting at $X (Y)$ is found to be very small ($\delta E\approx 0.3$ meV). Since the unit cell breaks four-fold symmetry in the $ab$-plane, $X$ and $Y$ directions in the Brillouin zone are inequivalent. This results in the states near $Y$ to be shifted away from the Fermi level to $-0.25$ eV.

As more layers are stacked along the $c$-axis, the band gap reduces converging to $1.719$ eV in the bulk structure.  Concomitantly, the valence band along $X-M$ becomes more dispersive, along with a decrease in the band splitting at $X$. This change in the curvature of the valence band appears to be driven by an enhancement in the S-$p_{z}$ / Ni-$d_{z^2}$ hybridized states concentrated around $X$. This mechanism, is similar to the one found in MoS$_2$, where the interfacial Mo-$d_{z^2}$ and S-$p_{z}$ character states facilitate the direct-indirect transition. \cite{trainer2017inter} Interestingly, we find significant $k_z$ dispersion of the band structure in the bulk crystal of NiPS$_3$ resulting from significant interlayer coupling. The abrupt suppression of long-range ordering upon thinning suggests interlayer coupling plays a key role stabilizing the various magnetic states.

The electron effective mass is a critical quantity in understanding carrier transport and designing new devices. In the semi-classical picture, the acceleration of an electron in an applied electric field is described by Newton's second law\cite{Ashcroft} and defines the inertial effective mass $m^*$ as being inversely proportional to the curvature of the energy dispersion at a particular point in the Brillouin zone,
\begin{align}
\frac{1}{m^*}=\frac{1}{\hbar^2}\left| \frac{\partial^2E(\mathbf{k})}{\partial \mathbf{k}^2} \right|.
\end{align}
This expression lends itself to straightforward evaluation from {\it ab initio} band structures. To estimate the effective masses of NiPS$_3$ for various thicknesses we used a three point forward finite difference equation, \cite{whalley2019impact,Whalley_JOSS2018}
\begin{align}
\frac{\partial^2E(\mathbf{k})}{\partial \mathbf{k}^2}=\frac{E(k_{2+i})-2E(k_{1+i})+E(k_{i})}{k_{i+1}-k_i},
\end{align}
to numerically evaluate the curvature at various high-symmetry points and band extrema along high-symmetry lines in the Brillouin zone. 

Table~\ref{table:effmass} gives the theoretically predicted effective masses for various thicknesses of NiPS$_3$ at different points in the Brillouin zone. Overall, the effective masses are significantly heavier than those found in Si, and other standard semiconductors, exhibiting masses of 9-10 $m_e$ on average. The masses of the valence and conduction bands at $\Gamma$ along $x$, $y$, and diagonal do not vary much with layer number. Whereas, interestingly, those at $Y$ along the diagonal ($y$-axis) appear to increase (decrease) for increased thicknesses. The remaining high-symmetry points exhibit masses that follow a non-monotonic progression with layer number. For the monolayer, bilayer, and trilayer, the valence band edge at each high-symmetry point is a saddle-point, exhibiting both electron and hole like curvatures. The bulk is more interesting, for which the band curvature at $X_y$, $M_y$, and $M-\Gamma$ display a strong $k_z$ dependence. For the conduction band, the $\Gamma$ point is distinctly electron like except when $k_z=\pi$ in the bulk phase. In contrast, $Y$ is hole like for all thicknesses and $k_z$ values in the bulk.

An important question in the physics of any material class is how their physical properties are linked to their electronic structure. A necessary step towards addressing this inquiry, is to be able to connect our theoretical ground state electronic structure to experimental measurements, that is, a manner by which to judge the quality of the theoretically obtained description. To this end, we calculate the dielectric tensor --a major ingredient in the interaction between light and matter-- and compare the results to experimental observations. The imaginary part of the dielectric tensor within the independent particle approximation is determined by a summation over empty states using the expression 
\begin{align}
\varepsilon^{(2)}_{\alpha\beta}(\omega)=&\frac{4\pi^2e^2}{\Omega}\lim\limits_{q \to 0} \frac{1}{q^2}\sum_{cv\mathbf{k}}2w_{\mathbf{k}}\delta\left(\varepsilon_{c\mathbf{k}}-\varepsilon_{v\mathbf{k}} -\omega \right) \\
&\times \braket{u_{c\mathbf{k}+\mathbf{e}_{\alpha}q}|u_{v\mathbf{k}}}  \braket{u_{c\mathbf{k}+\mathbf{e}_{\beta}q}|u_{v\mathbf{k}}}^{*} \nonumber
\end{align}
where the indices $c$ and $v$ refer to the conduction and valance band states, respectively, $u_{c\mathbf{k}}$ is the cell periodic part of the orbitals at crystal momentum $\mathbf{k}$ in the irreducible Brillouin zone of weight $w_{\mathbf{k}}$, and the vectors $\mathbf{e}_{\alpha}$ are unit vectors for the three Cartesian directions.\footnote{Since we are employing a Meta-GGA density functional, the derivative of the cell-periodic part of the orbitals was computed using the finite difference scheme as implemented in vasp. } The real part of the dielectric tensor $\varepsilon^{(1)}_{\alpha\beta}$ is obtained by the usual Kramers-Kronig transformation
\begin{align}
\varepsilon^{(1)}_{\alpha\beta}(\omega)=1+\frac{2}{\pi}\mathcal{P}\int_{0}^{\infty} \frac{\varepsilon^{(2)}_{\alpha\beta}(\omega^{\prime})\omega^{\prime}}{\omega^{\prime 2}-\omega^{2}}d\omega^{\prime}
\end{align}
where $\mathcal{P}$ denotes the principal value.

\begin{figure}[ht!]
\includegraphics[width=0.5\textwidth]{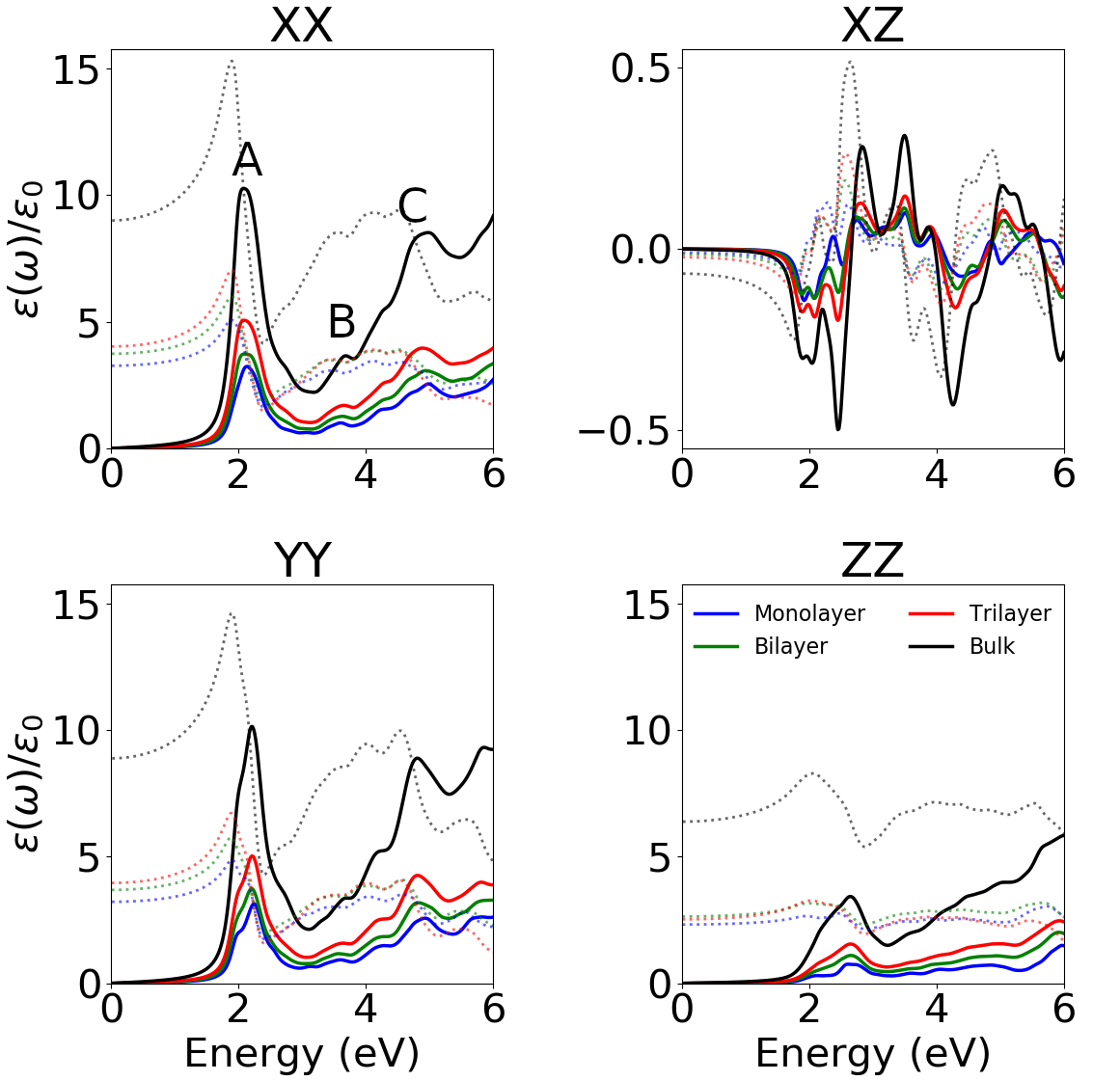}
\caption{(color online)  The non-zero real (dashed lines) and imaginary (solid lines) components of the dielectric tensor of NiPS$_3$ in the Zig-Zag magnetic phase for the monoclinic bulk, monolayer, bilayer, and trilayer structures. } 
\label{fig:OPTICAL}
\end{figure}

Figure~\ref{fig:OPTICAL} shows the non-zero real (dashed lines) and imaginary (solid lines) components of the dielectric tensor for various thicknesses of NiPS$_3$. Two main transitions are distinctly seen at approximately 2 eV and 5 eV, along with a small peak at 3.5 eV in the imaginary part of the $xx$ and $yy$ tensor components for all thicknesses.  Due to the layered nature of NiPS$_3$, the amplitude of the leading transition in $\varepsilon^{(2)}_{zz}$ is less than half of the corresponding peak in the in-plane tensor elements and is blue shifted by $0.5$ eV. Combining the diagonal components on average, we find the $A$, $B$, and $C$ labeled peaks in good accord with the spectroscopic ellipsometry report by Kim {\it et al}. \cite{kim2018charge} Comparing to Fig.~\ref{fig:BANDSDOS}, transition $A$ is produced by promoting an electron from the valance to conduction band edges along $X-M $. The higher energetic transitions $B$ and $C$ originate from bands $\sim 1.0$ eV below the Fermi level connecting to the flat conduction bands along $\Gamma-Y$ and $X-M$.  Furthermore, our theoretically predicted electronic band gap is in very good agreement with the leading edge of the optical conductivity of Ref.~\onlinecite{kim2018charge}. Upon thinning NiPS$_3$ from the bulk, the leading edge peak blues shifts by $0.08$ eV, inline with the corresponding increase in the band gap. In $\varepsilon^{(2)}_{zz}$ a slight shoulder appears, mainly driven by the change in energy and curvature of the bands along $X-M$.

Interestingly, it was shown experimentally \cite{kim2018charge} that, as the sample is heated above $T_N$, the position of the $A$, $B$, and $C$ peaks display no strong reorganization. Moreover, only a gradual decrease in peak amplitude is observed as a function of temperature. This suggests two possible mechanisms underling this behavior: (i) the interlayer coupling is highly dependent on temperature, or (ii) the presence of strong electronic correlations. For the former case, if the $c$-axis lattice parameter were to significantly increase proportionally with temperature, 
the various layers may decouple, making each layer behave electronically independent at high temperatures. However, the $c$ lattice parameter only changes by 0.5\%, ruling out the first scenario.\cite{wildes2015magnetic} 

In our analysis of the frequency dependent dielectric function, we find the three predominant peaks to be a direct consequence of the underlying Zig-Zag magnetic order, since they are formed by the opening of the electronic band gap. Therefore, following scenario (ii), we expect that as the temperature is increased from below to above $T_N$ the gap will persist, due to short-range order driven by strong electronic correlations. This behavior follows the experimental results. Moreover, since the energetic separation between the Zig-Zag and N\'eel type magnetic orders in the monolayer is  marginal, we may consider the monolayer case to mimic the short-range order phase. Comparing the theoretical dielectric function of the monolayer to the experimental spectra for high temperatures, we find the monolayer to simulate the decrease in peak amplitude.  Therefore, our results suggest that the gap in NiPS$_3$ appears to be driven by strong electronic correlations of the Mott-Hubbard type, rather than the Slater-Stoner type, and for $T>T_N$ we find the presence of a pesudogap regime similar to the cuprate high-temperature superconductors.

When a linearly polarized light is reflected from a magnetic material, the reflected light is typically elliptically polarized. The angle through which the ellipse rotates is called the Kerr rotation angle. This phenomenon is  exceptionally useful to mark structural phase transitions\cite{tsvetkov2002optical} and to give direct insight into the local, microscopic magnetism in condensed matter systems.~\cite{erskine1975calculation,reim1990ferromagnetic} Since NiPS$_3$ exhibits a non-zero $\varepsilon_{xz}$ tensor component an appreciable Kerr angle is expected. Here, we track the changes in the Kerr parameters with layer number to see if they can be used as an indicator of film-thickness.

\begin{figure}[ht!]
\includegraphics[width=0.90\columnwidth]{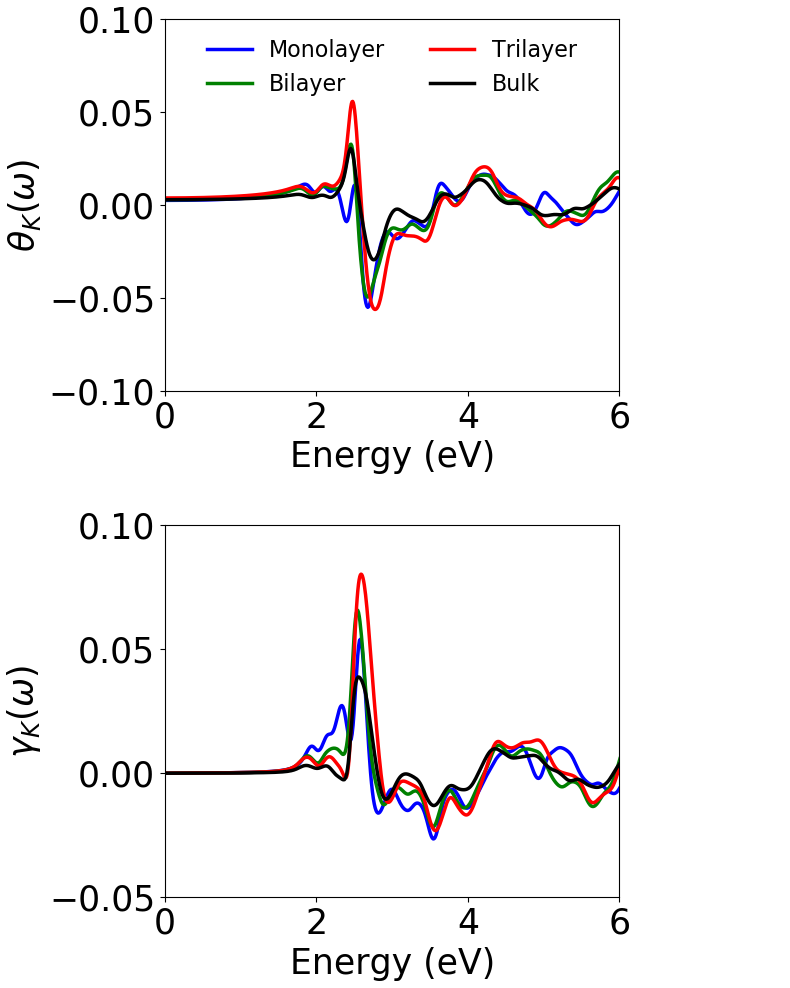}
\caption{(color online) The Kerr parameters $\theta_K$ and $\gamma_K$ as a function of energy for various thicknesses of NiPS$_3$.} 
\label{fig:KeerAngle}
\end{figure}

The Kerr parameters are related to the dielectric function by the following, 
\begin{align}\label{eq:kerr}
\Psi_{k}=\theta_{K}+i\gamma_{K}=\frac{-\varepsilon_{xz}}{(\varepsilon_{xx}-1)\sqrt{\varepsilon_{xx}}}
\end{align}
for a polar geometry in the small angle limit. Here, the photon propagates along the $y$-direction and describes a linearly polarized wave with the electric field  along the $x$-direction. \cite{erskine1975calculation,sangalli2012pseudopotential} The off diagonal dielectric tensor element $\varepsilon^{(2)}_{xz}$ [Fig.~\ref{fig:OPTICAL}] is non-zero and oscillates about zero following a clear amplitude enhancement with increased thickness, until saturation in the bulk.

\begin{figure}[ht!]
\includegraphics[width=1.0\columnwidth]{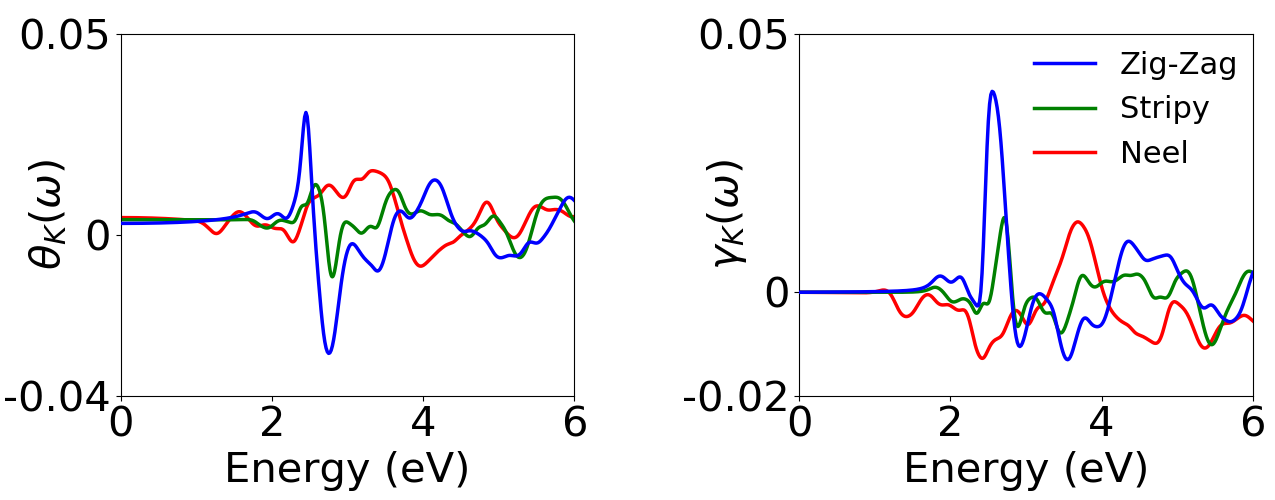}
\caption{(color online) The Kerr parameters $\theta_K$ and $\gamma_K$ as a function of energy for the Zig-Zag, N\'eel, and Stripy magnetic configurations for bulk NiPS$_3$ in the Zig-Zag AFM phase. } 
\label{fig:KERRMAG}
\end{figure}

Figure~\ref{fig:KeerAngle} shows the complex Kerr parameters from Eq.~(\ref{eq:kerr}) as a function of energy for various thicknesses of NiPS$_3$. The values are small $\sim 0.1$, as expected for optical wavelengths.\cite{erskine1975calculation} Interestingly, the leading edge of $\gamma_{K}$ near $2$ eV follows a regular progression with layer number, where the spectra starts with a positive value and gradually reduces, until flipping sign for  a film thickness larger than two layers. $\theta_{K}$ exhibits a similar behavior limiting to zero in the bulk case. Furthermore, Fig.~\ref{fig:KERRMAG} shows the Kerr parameters in the bulk for the Zig-Zag, N\'eel, and Stripy magnetic configurations. The Zig-Zag order produces the largest Kerr response, while the Stripy order produces a similar but weaker values. Interestingly, the N\'eel order produces a very different energy dependence of $\theta_{K}$ and $\gamma_{K}$, shifting the main peak at $4$ eV (See the Appendix for the dielectric fuctions in the Zig-Zag, N\'eel, and Stripy phases.). Therefore $\theta_{K}$ and $\gamma_{K}$, though small, might hold promise as an optical descriptor of magnetic arrangement and film thickness in NiPS$_3$ and other magnetic 2D materials.

\section{Effect of an External Static Electric Field}\label{sec:efield}

Voltage control of electronic band gaps and magnetism has been intensely pursued during the past few decades 
\cite{song2017recent,bader2010spintronics,chappert2010emergence,wolf2001spintronics,vzutic2004spintronics,chang2014thickness,sun2010electrically,niranjan2010electric} not only due to the direct connection to application in the  miniaturization of magneto-electronics needed for spintronics and quantum information technologies, but also for the rich fundamental physics at the heart of the interplay between charge, spin, orbital, and lattice degrees of freedom.

\begin{figure}[ht!]
\includegraphics[width=\columnwidth]{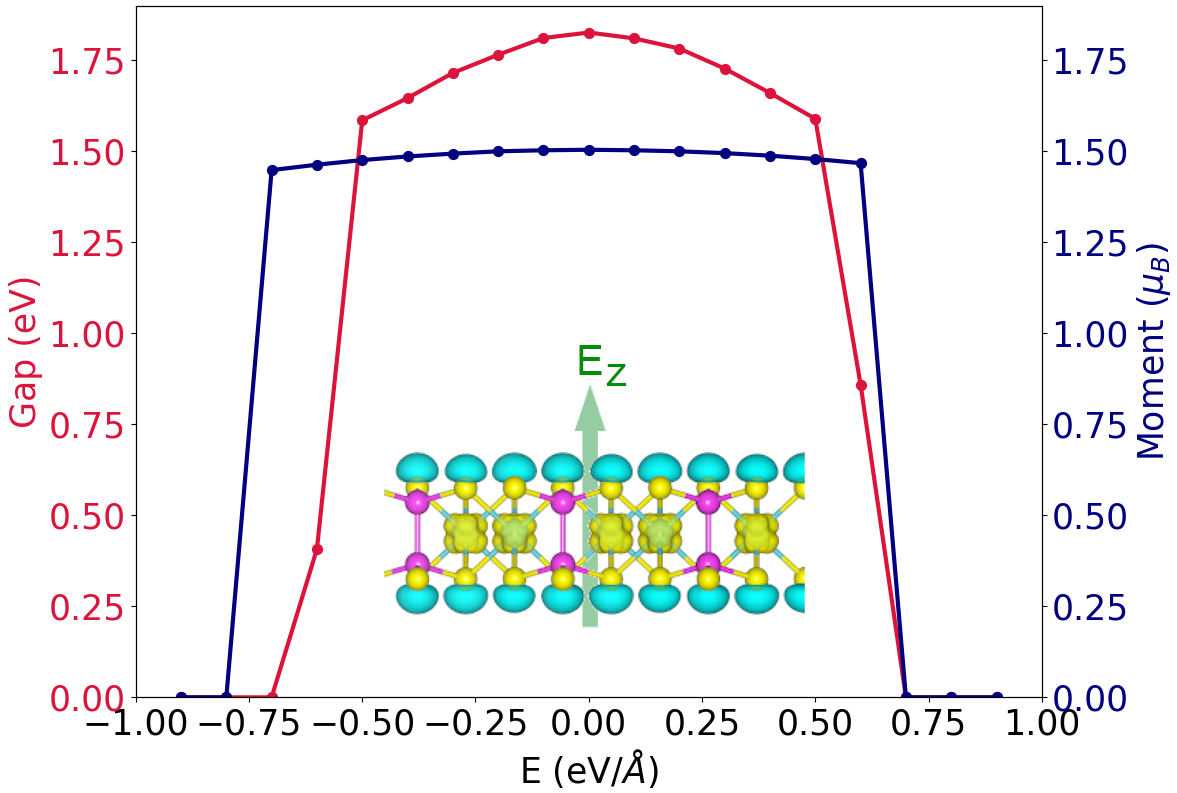}
\caption{(color online) Evolution of the electronic band gap (crimson line) and the nickel magnetic moment (blue line) for monolayer NiPS$_3$ with external applied electric field. $\mathbf{E}$ is taken positive along $z$-direction, and is shown schematically with the crystal structure of monolayer NiPS$_3$.   } 
\label{fig:Efield}
\end{figure}

Figure \ref{fig:Efield} shows the magnitude of the nickel magnetic moment for monolayer NiPS$_3$ as a function of electric field positively aligned along the $z$-axis. As the electric field strength increases the magnetic moments decrease. At  an $E_{z}$ of $0.6 ~(-0.7)$ eV/\AA~a first-order transition occurs, quenching the magnetic moment. Additionally, the maximum deviation of the magnetic moments before collapse is only 3\%, allowing for clean high-to-low and low-to-high transitions needed in magnetic based switches. Simultaneously, the electronic band gap decreases with applied field faster than that of the magnetic moments, reaching a max 13\% deviation before passing through an insulator-metal transition at a field strength of $\pm 0.6$ eV/\AA. 

Figure~\ref{fig:Efield} (inset) shows the charge density difference between the insulating-magnetic and metallic-non-magnetic phases where the positive (negative) charge clouds are colored yellow (light blue). A clear migration of charge from the sulfur atoms to the nickel atomic sites is observed, with the iso-level surface resembling a linear combination of $d_{x^2-y^2}$ and $d_{z^2}$ orbitals. Phenomenologically, this can be rationalized as follows: The sulfur atoms produce an octahedral crystal field splitting the spherically symmetric manifold of nickel $d$-orbitals into $t_{2g}$ and $e_{g}$ levels.  Since Ni carries a $2^{+}$ ionic state, the degenerate $d_{x^2-y^2}$ and $d_{z^2}$ orbitals comprising the $e_{g}$ subspace are half-filled.  Due to significant on-site electron repulsion in the 3$d$ transition metals, the spin degeneracy of the  $e_{g}$ orbitals is split to lower the total energy of the system, producing a Zig-Zag AFM order across nickel atomic sites. When the electric field is applied, the system reorganizes to screen the external field, pushing the system towards metallicity -- as indicated by the quickly reduced band gap -- which in turn reduces the effect of the on-site Coulomb potential. Once the on-site Coulomb potential is sufficiently screened, the AFM order collapses. This is similar to the mechanism of manipulating magnetism in platinum and iron as discussed in Refs.~\onlinecite{rana2016electric,weisheit2007electric}.

%% Conclusions

\section{Concluding Remarks}\label{sec:conclusion}
By examining the ground state electronic structure of NiPS$_3$ as a function of the film thickness, we tracked the systematic changes in the electronic and magentic structure. Key properties, including the competition of magnetic ordering states in the monolayer and the persistnce of the experimental electric gap resemble the characterize features of the pseudogap in cuprate high-temperature superconductors. In order to further elucidate the connection between NiPS$_3$ and the cuprates, if any, further doping studies are needed to examine the evolution of the electronic and magnetic phases, and compare to the phenomenology of the cuprates. Additionally, the layer degrees of freedom gives us another knob, by which to tune the ground state, providing a platform to study the competition of various short-range magnetic and charge fluctuations.

%% Acknowledgments

\begin{acknowledgments}
This work was carried out under the auspices of the U.S. Department of Energy (DOE) National Nuclear Security Administration under Contract No. 89233218CNA000001.
It was supported by the LANL LDRD Program, and in part by the Center for Integrated Nanotechnologies, a DOE BES user facility, in partnership with the LANL Institutional Computing Program for computational resources. 
\end{acknowledgments}

\appendix*

\section{Dielectric Tensor for Various Magnetic Configurations}\label{A:optical}

\begin{figure}[ht!]
\includegraphics[width=1.01\columnwidth]{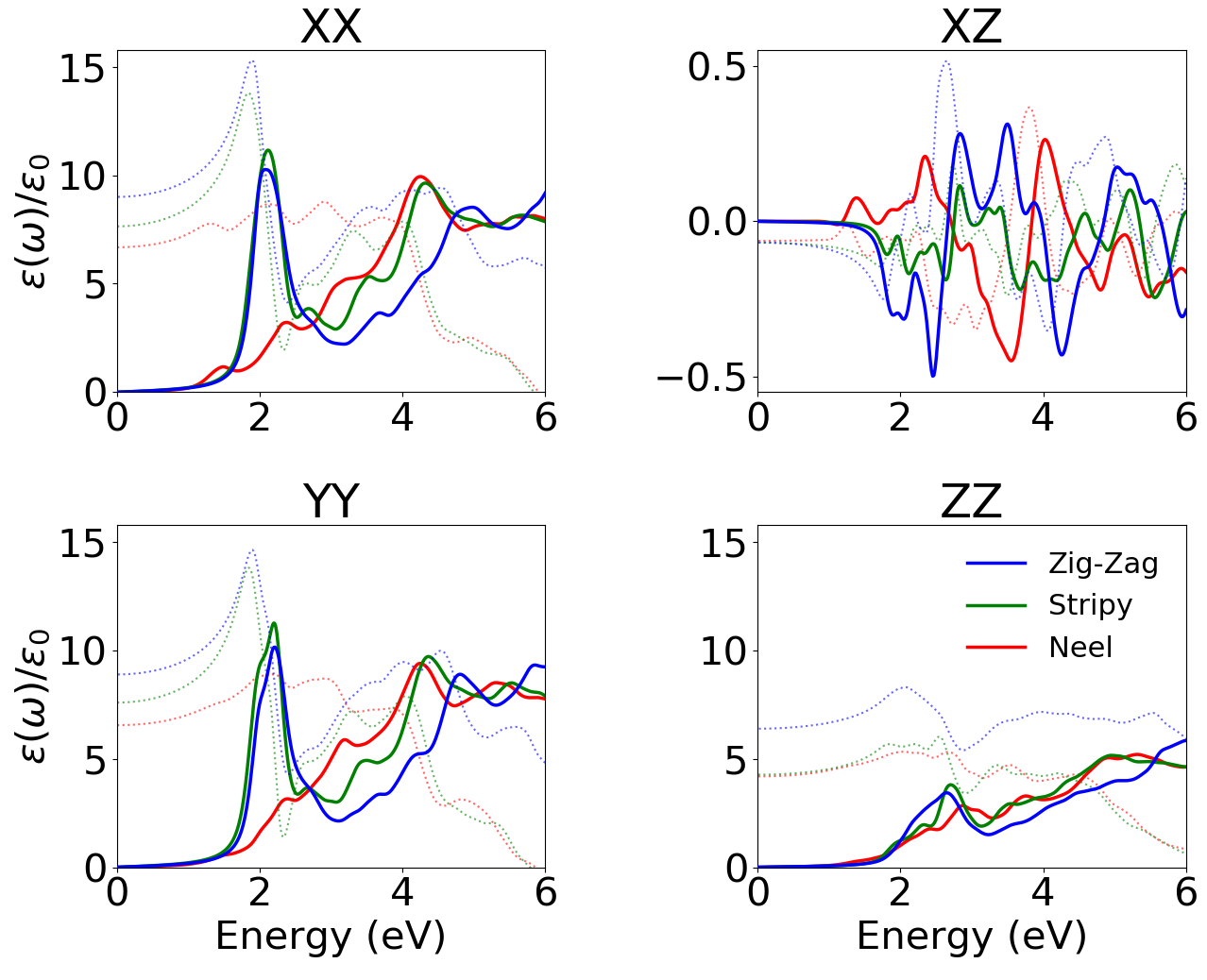}
\caption{(color online)  The non-zero real (dashed lines) and imaginary (solid lines) components of the dielectric tensor of NiPS$_3$ in the bulk monoclinic structure for the Zig-Zag, N\'eel, and stripy magnetic orders. } 
\label{fig:DTmag}
\end{figure}

Figure~\ref{fig:DTmag} compares the energy dependent dielectric tensor for the Zig-Zag, N\'eel, and Stripy magnetic configurations in the bulk monoclinic structure of NiPS$_3$. The dominant transitions in the Stripy are very similar to those of the Zig-Zag arrangement, except for slight red sifts in the $B$ and $C$ transitions. Interestingly, the N\'eel magnetic arrangement does not display a divergent transition at $\approx 2$ eV as compared to the other case.

\bibliography{NiPS3_Refs}

\end{document}